\documentclass[twocolumn,prl,showpacs,amsmath,amssymb]{revtex4}

\usepackage{graphicx}

\DeclareMathOperator*{\argmin}{argmin}

\newcommand{\Exp}[1]{\left\langle #1 \right\rangle}

\newcommand{\st}{s}

\newcommand{\m}{m}
\newcommand{\bxi}{\xi}

\newcommand{\where}{\,|\,}
\newcommand{\Prb}{\mathbb{P}}
\newcommand{\arctanh}{\mathrm{arctanh\,}}

\newcommand{\abs}[1]{\left\lvert #1 \right\rvert}
\newcommand{\N}[1]{\mathrm{#1}}
\newcommand{\nel}[1]{\abs{#1}}

\begin{document}

\preprint{cond-mat/0408378}

\title{Spin-glass phase transitions on real-world graphs}

\author{J.M.\ Mooij}
 \email{j.mooij@science.ru.nl}
\author{H.J.\ Kappen}
 \email{b.kappen@science.ru.nl}
\affiliation{Dept.\ of Biophysics, Inst.\ for Neuroscience, Radboud University Nijmegen, Geert Grooteplein Noord 21, 6525 EZ Nijmegen, The Netherlands}

\date{September 16, 2004}

\begin{abstract}  
We use the Bethe approximation to calculate the critical temperature for the
transition from a paramagnetic to a glassy phase in spin-glass models on
real-world graphs. Our criterion is based on the marginal stability of the
minimum of the Bethe free energy.  For uniform degree random graphs (equivalent
to the Viana-Bray model) our numerical results, obtained by averaging single
problem instances, are in agreement with the known critical temperature
obtained by use of the replica method. Contrary to the replica method, our
method immediately generalizes to arbitrary (random) graphs. We present new
results for B\'arabasi-Albert scale-free random graphs, for which no analytical
results are known. We investigate the scaling behavior of the critical
temperature with graph size for both the finite and the infinite connectivity
limit. We compare these with the naive Mean Field results. We observe that the
Belief Propagation algorithm converges only in the paramagnetic regime.
\end{abstract}

\maketitle
Sparse networks with non-uniform topology occur in many contexts ranging from
biology and sociology to communication systems and computer science
\cite{AlbertBarabasi02}. Prominent examples of mathematical models designed to
capture the topological features of these so-called ``real-world'' graphs are
uniform degree random graphs \cite{ErdosRenyi59}, scale-free networks
\cite{BarabasiAlbert99} and small-world graphs \cite{WattsStrogatz98}. While
much research has been done on topological properties of these complex
networks, not much seems to be known about the properties of such networks when
nodes contain dynamic variables and links represent positive
or negative interaction between them. Examples of complex networks with
interacting nodes are networks of neurons in the brain \cite{Elegans},
communication in networks of sensors \cite{PaskinGuestrin04}, combinatoric
optimization problems \cite{MezardZecchina02} and protein folding
\cite{Wuchty01,KussellShakhnovich02}. In this Letter, we take a statistical
physics approach to study the behavior of such systems.

We study generalizations of the SK model \cite{SherringtonKirkpatrick75} where
we replace the fully connected underlying graph by complex random graphs. We
use the Bethe approximation, an extension of the naive Mean Field (MF)
approximation, to calculate the critical temperature corresponding to the phase
transition from a paramagnetic to a spin-glass-like phase. The transition is
characterized by a marginal instability of the Hessian of the Bethe free
energy.  For uniform degree random graphs, averaging over the interactions and
the underlying graph instances yields results consistent with known analytical
results obtained by use of the replica method \cite{VianaBray85}. Contrary to
the replica method, our method immediately generalizes to arbitrary (random)
graphs. We present results for scale-free networks, for which no
analytical results are known to the best of our knowledge (however, for the
ferromagnetic case on random graphs with arbitrary degree distribution, see
\cite{LeoneVazquezVespignaniZecchina02}). 

Another advantage of the Bethe approximation is that it yields results for
finite $N$ and for specific problem instances of the interactions. This is
important for technologically relevant applications \emph{viz.} image
restoration \cite{Tanaka02}, artificial vision
\cite{FreemanPasztorCarmichael00}, decoding of error-correcting codes
\cite{McElieceMacKayCheng98} and medical diagnosis \cite{Kappen02}. The
problems in these applications can generally be reformulated in terms of 
thermodynamic systems defined on graphs, and solving them amounts to the
determination of the Boltzmann distribution for these systems.

Let $G=(V,B)$ be an undirected labelled graph without self-connections, defined
by a set of vertices $V=\{1, \dots, N\}$ (corresponding to the spins) and a set
of edges $B \subseteq \{(i,j) \,|\, 1 \le i < j \le N\}$ (corresponding to
non-zero interactions between the spins). We define the \emph{adjacency matrix}
$M$ corresponding to $G$ as follows: $M_{ij} = 1$ if $(ij) \in B$ or $(ji) \in
B$ and 0 otherwise. Denote by $N_i$ the set of neighbors of vertex $i$, and
denote the degree (connectivity) of vertex $i$ by $d_i := \nel{N_i} =
\sum_{j\in V} M_{ij}$. We define the \emph{average degree} $d := \tfrac{1}{N}
\sum_{i\in V} d_i$ and the \emph{maximum degree} $\Delta := \max_{i\in V} d_i$.

To each vertex $i\in V$ we associate an Ising spin $s_i$, taking values in
$\{-1,+1\}$. Let $J$ be a symmetric $N\times N$ matrix representing the
interactions, which we do not further specify for the moment, except for the
constraint that $J$ should be compatible with the adjacency matrix $M$, i.e.\
$J_{ij} = 0$ if $M_{ij} = 0$. For the Hamiltonian of the system we take 
  $$H = -\sum_{(i,j) \in B} J_{ij} s_i s_j$$ 
and we study the corresponding Boltzmann
distribution over the configurations $\st = (s_1,\dots, s_N) \in \{-1,+1\}^N$:
  $$\Prb(\st) = \frac{1}{Z} \exp \left(\beta \sum_{(i,j) \in B} J_{ij} s_i s_j\right).$$ 
Note that, because of the sign reversal symmetry, the exact
magnetizations are given by $\Exp{s_i} = 0$.

The Bethe free energy \cite{YedidiaFreemanWeiss01} can
be written \cite{WellingTeh01} as a function of the parameters $\m =
(m_1,\dots,m_N)$ and $\bxi = \{\xi_{ij}\}_{(ij)\in B}$:
  \begin{multline*}
  F_{Be}(\m,\bxi) := -\beta\sum_{(ij) \in B} J_{ij} \xi_{ij} \\
  + \sum_{i=1}^N (1-d_i) \sum_{s_i=\pm 1} \eta\left(\frac{1+m_i s_i}{2}\right) \\
  + \sum_{(ij)\in B} \sum_{s_i,s_j=\pm 1}
    \eta\left(\frac{1+m_i s_i+m_j s_j + s_i s_j \xi_{ij}}{4}\right)
  \end{multline*}
where $\eta(x) := x \log x$.
The Bethe approximation consists of minimizing the Bethe free energy
with respect to the parameters $\m$ and $\bxi$ under the following constraints:
  \begin{align*}
  & -1 \le m_i \le 1 \\
  & -1 \le \xi_{ij} \le 1\\
  & 1+m_i \sigma+m_j \sigma' + \xi_{ij}\sigma \sigma'  \ge 0 \quad \text{for all $\sigma, \sigma' = \pm 1$.}
  \end{align*}
The parameters $m_i$ are then approximations for the means $\langle s_i
\rangle$, while the $\xi_{ij}$ are approximations for the moments $\langle s_i
s_j \rangle$ for $(ij) \in B$; if $G$ is a tree, these approximations are exact
\cite{Baxter82}.
It is straightforward to check that $m_i = 0$ and $\xi_{ij} = \tanh
(\beta J_{ij})$ is a stationary point of $F_{Be}$, which we will call the
\emph{paramagnetic} solution. For this solution to be a minimum, the Hessian of
$F_{Be}$ has to be positive definite. Defining $\xi_{ji} := \xi_{ij}$ for
$(i,j) \in B$ and $\xi_{ij} := 0$ if $(i,j) \not\in B$ and $(j,i) \not\in B$, this is
equivalent to the matrix 
  \begin{equation}\begin{split}
  \label{eq_A_Be}
  (A_{Be})_{ik} & := \frac{\partial^2 F_{Be}}{\partial m_k \partial m_i}\Big|_{m=0,\xi=\tanh (\beta J)} = \\ 
  & \delta_{ik} \left( 1 + \sum_{j\in N_i} \frac{\xi_{ij}^2}{1-\xi_{ij}^2}\right)
  - \frac{\xi_{ik}}{1-\xi_{ik}^2}
  \end{split}\end{equation}
being positive definite, since the $\partial F_{Be} / \partial m_k \partial
\xi_{ij}$ part vanishes and the other block $\partial F_{Be} / \partial
\xi_{ij} \partial \xi_{kl}$ is positive definite for all possible parameter
values. 

We will compare the results with the naive Mean Field approximation. The
MF free energy is given by
  $$F_{MF}(\m) := -\beta\sum_{(i,j) \in B} J_{ij} m_i m_j + \sum_{i=1}^N \sum_{s_i=\pm 1} \eta \left(\frac{1+m_i s_i}{2}\right)$$
and has to be minimized. The Hessian is simply
  \begin{equation}\label{eq_A_MF}
  (A_{MF})_{ij} := \frac{\partial^2 F_{MF}}{\partial m_i \partial m_j} = -\beta J_{ij} + \delta_{ij} \frac{1}{1-m_i^2}.
  \end{equation}
Note that $\m = 0$ is a stationary point of the MF free energy---not
necessarily a minimum---which we also call the \emph{paramagnetic} solution.

The stability of the
paramagnetic MF solution implies the stability of the paramagnetic
Bethe solution, as we shall now demonstrate. Define $A := A_{Be} - A_{MF}$; we
have to show that $A$ is positive semidefinite. Substituting \eqref{eq_A_Be}, \eqref{eq_A_MF} and the relation $\xi_{ij} = \tanh (\beta J_{ij})$:
  $$A_{ij} =  - \frac{\xi_{ij}}{1 - \xi_{ij}^2} + \arctanh \xi_{ij} + \delta_{ij} \sum_{k=1}^N \frac{\xi_{ik}^2}{1 - \xi_{ik}^2}.$$
Note that $A$ is symmetric and that all its diagonal entries are nonnegative.
Furthermore, $A$ is diagonally dominant, i.e.\ for all $i=1,\dots,N$:
  $$A_{ii} - \sum_{\substack{j=1 \\ j \ne i}}^N \abs{A_{ij}},
  \ge 0$$
which immediately follows from the following inequality:
  $$\frac{x^2}{1-x^2} - \abs{\arctanh x - \frac{x}{1-x^2}} \ge 0$$
that holds for all $x \in (-1,1)$.
A simple application of Gerschgorin's Circle Theorem yields the desired 
result.

Before we embark on spin-glasses, let us first shortly discuss the purely 
ferromagnetic case, where we take all
nonzero interactions to be equal and positive, i.e.\ $J = M$. Both
approximations are stable for high temperature (i.e.\ small $\beta$) but break
down at some critical $\beta_c$, where the Hessians develop negative 
eigenvalues and the unique minimum of the free energy splits into two 
minima.
For MF, this critical $\beta_c^{MF}$ is given by $\beta_c^{MF} = 1/\lambda_1,$
where $\lambda_1$ is the principal eigenvalue of the adjacency matrix $M$, as
easily follows from \eqref{eq_A_MF}. Since $d \le \lambda_1 \le \Delta$, we 
immediately get the following bound on the ferromagnetic MF critical 
temperature:
  $$\frac{1}{\Delta} \le \beta_c^{MF} \le \frac{1}{d}.$$
For the Bethe approximation, the critical value $\beta_c^{Be}$ at which 
$A_{Be}$ develops a negative eigenvalue can be shown to satisfy the following
bound:
  $$\frac{1}{\Delta - 1} \le \frac{1}{\lambda_1 - 1} \le \tanh \beta_c^{Be} \le \frac{1}{d-1}.$$
In practice, for the graph topologies that we investigated, the critical
values $\beta_c^{MF}$ and $\beta_c^{Be}$ differ only slightly. This is not the
case for spin-glass like interactions, where the Bethe approximation 
clearly outperforms the MF approximation, as we shall discuss shortly.

\begin{figure}
\includegraphics[width=8.6cm]{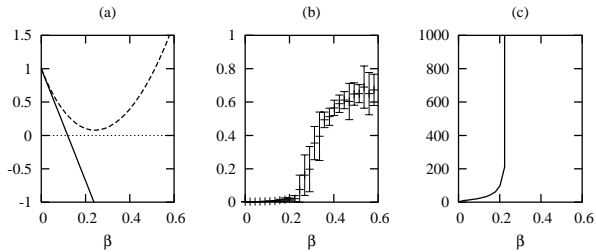}
\caption{\label{fig:sg_transition}Transition from paramagnetic to spin-glass
phase for an ER graph with $N=100$ and $d=20$ and Gaussian interactions. From
left to right: (a) minimal eigenvalues of $A_{MF}$ (solid) and of $A_{Be}$
(dashed); (b) Edwards-Anderson order parameter $q_{EA}$; (c) number of Belief
Propagation iterations before convergence.}
\end{figure}

In the following, we take the nonzero interactions $\{J_{ij}
\where M_{ij}=1, i>j\}$ to be independent Gaussian random variables with mean
$0$ and variance $1$. We first discuss the typical single-instance behavior.
Fig.\ \ref{fig:sg_transition}(a) shows how the minimal eigenvalues of the
stability matrices $A_{MF}$ and $A_{Be}$ typically depend on the inverse
temperature $\beta$. Varying the graph topology only scales the axes (except
for tree-like or even sparser graphs), but qualitatively the picture remains
the same. For the Mean Field method, the paramagnetic solution becomes unstable
for $\beta > \beta_c^{MF} := \sup \{\beta > 0; A_{MF} > 0 \}.$ 
For the Bethe approximation however, we typically find that with increasing
$\beta$, the smallest eigenvalue $\lambda_{\N{min}}(A_{Be})$ first decreases 
until it becomes approximately zero at some critical $\beta_c^{Be}$, after 
which it starts increasing again. At this critical $\beta_c^{Be}$, which we 
define as
  $$\beta_c^{Be} := \argmin_{\beta>0} \lambda_{\N{min}}(A_{Be}),$$
the paramagnetic solution is \emph{marginally} stable. Our numerical experiments
show that in the thermodynamic limit $N\to\infty$, the minimal eigenvalue 
$\lambda_{\N{min}}(A_{Be})$ evaluated at $\beta_c^{Be}$ converges to 0.
Furthermore, Monte Carlo simulations using the Metropolis
algorithm show that the Edwards-Anderson parameter $q_{EA} := \frac{1}{N}
\sum_{i=1}^N \Exp{s_i}^2$ becomes positive at the transition (see Fig.\
\ref{fig:sg_transition}(b)). We therefore interpret this marginal instability 
as a phase transition to a spin-glass like state
\cite{MezardParisiVirasoro87}.

\begin{figure*}
\includegraphics{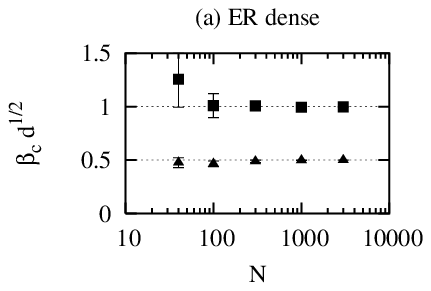}
\includegraphics{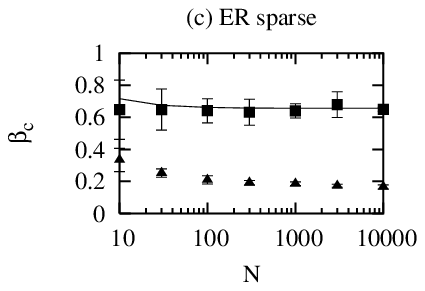}
\includegraphics{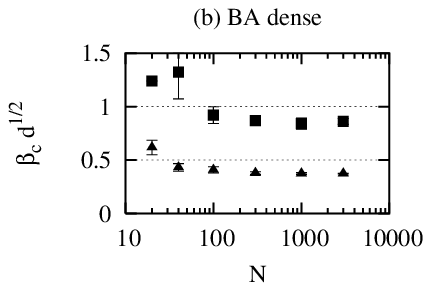}
\includegraphics{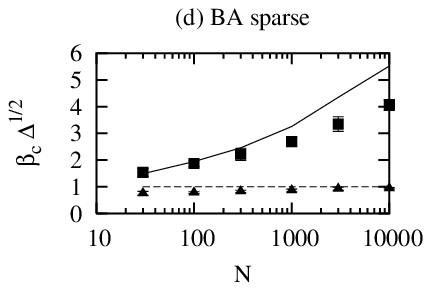}
\caption{\label{fig:ndep}
Critical inverse temperatures $\beta_c^{MF}$ (MF approximation, triangles)
and $\beta_c^{Be}$ (Bethe approximation, squares) for the spin-glass transition
vs.\ graph size $N$, averaged over interactions and over graph instances.  From left to right: 
(a) ER, dense limit, $p = 0.1$ (the vertical axis is rescaled by $1 /
\sqrt{d}$); 
(b) ER, sparse limit, $d = 4$; the solid line shows $\beta_c^{VB}$ 
according to eqn.\ (\ref{eq:replica_pa_sg}); 
(c) BA, dense limit, $p = 0.1$ (the vertical axis is rescaled by $1 /
\sqrt{d}$); 
(d) BA, sparse limit, $d = 10$ (the vertical axis is rescaled by $1 /
\sqrt{\Delta}$); the solid line is $1/\sqrt{d}$ and the dashed line is
$1/\sqrt{\Delta}$.}
\end{figure*}

In the remainder of this article we study the dependence of the Bethe and MF
critical temperatures on the graph topology and its size $N$.  Until now,
everything is valid for single instances (consisting of a specific choice of
$G$ and of $J$), which is obviously very useful in applications. In the
following we will average over graph instances and over interactions in order
to get an average-case analysis. 
We start with the uniform degree random graphs introduced and studied by Erd\H
os and R\'enyi \cite{ErdosRenyi59} (ER graphs in short). The ensemble consists
of graphs with $N$ vertices, where each pair of vertices is independently
connected with probability $p$. The degree distribution is approximately
Poisson for large $N$ and the expected average degree is $\Exp d = p (N-1)$.
The resulting model is also known in statistical physics as the Viana-Bray 
model of diluted spin-glasses \cite{VianaBray85}. Numerical results are shown 
in Fig.\ \ref{fig:ndep}.a and \ref{fig:ndep}.b.

For $N\to\infty$, we can state rigorous results about the scaling behavior of
the MF critical temperature. There are (at least) two ways to take the limit $N
\to\infty$, which result in different scaling behavior: the \emph{sparse} (or
finite connectivity) limit in which we fix the average degree $d$ and let
$N\to\infty$, and the \emph{dense} (infinite connectivity) limit in which we
fix $p$ and take $N\to\infty$. In the dense limit (Fig.\ \ref{fig:ndep}.a),
$\beta_c^{MF} \sim 1 / 2 \sqrt{d}$, which follows from a result of Khorunzhy
\cite{Khorunzhy01} stating that the principal eigenvalue of $J$ is
asymptotically equal to $2\sqrt{Np}$, if $d(N) \gg \log N$. On the other hand,
for $d(N) \ll \log N$, the spectral norm of $J$ is asymptotically much larger
than $\sqrt{Np}$, which implies that the MF $\beta_c^{MF}$ converges to zero in
the sparse limit (c.f.\ Fig.\ \ref{fig:ndep}.b).

For the Bethe approximation, we were not able to obtain rigorous results about
the scaling behavior of $\beta_c^{Be}$, but our numerical results turn out to
be in agreement with the known phase boundary derived using the replica method. 
In \cite{VianaBray85}, the following equation is derived for the critical
inverse temperature $\beta_c^{VB}$ corresponding to paramagnetic--spin-glass phase transition:
  \begin{equation}\label{eq:replica_pa_sg}
  \int \tanh^2 (\beta_c^{VB} x) \,dP(x) = \frac{1}{d},  
  \end{equation}
where $P$ is the probability distribution of the weights, which is Gaussian 
with mean 0 and variance 1 in our case. For large $d$, this yields a critical
inverse temperature of approximately $\beta_c^{VB} \approx 1 / \sqrt{d}$. As 
illustrated in Fig.\ \ref{fig:ndep}.a and Fig.\ \ref{fig:ndep}.b, the average 
Bethe critical $\beta_c^{Be}$ perfectly agrees with the value 
$\beta_c^{VB}$ obtained by Viana and Bray, in the dense as well as in the sparse
limit. This confirms our interpretation of the marginal instability as
a phase transition to a spin-glass like phase. In sharp contrast with MF, the 
Bethe critical $\beta_c^{Be}$ converges to a positive constant in the sparse
limit. In the dense limit the Bethe $\beta_c^{Be}$ is twice as large
as the MF $\beta_c^{MF}$.

We now present numerical results on Barab\'asi-Albert scale-free networks, for
which no results have been published before as far as we know. A phenomenon
often observed in real-world networks is that the degree distribution behaves
like a power law \cite{AlbertBarabasi02}, i.e.\ the number of vertices with
degree $\delta$ is proportional to $\delta^{-\alpha}$ for some $\alpha > 0$.
The first and intensely studied random graph model showing this behavior is due
to Barab\'asi and Albert \cite{BarabasiAlbert99}. 
The degree distribution has a power-law dependence
for $N\to\infty$: the probability that a randomly chosen vertex has a
particular degree $\delta$ is proportional to $\delta^{-3}$. The difference
between the maximum degree $\Delta$ and the average degree $d$ becomes quite
large for the BA model compared with ER random graphs. A natural question to ask
is what quantity will govern the spin-glass transition: the average degree $d$
or the maximum degree $\Delta$.

In the dense limit (see Fig.\ \ref{fig:ndep}.c), we find $\beta_c^{MF} \propto
1/2\sqrt{d}$ and $\beta_c^{Be} \propto 1/\sqrt{d}$. This is very similar to the
ER case, up to a constant factor of order 1. Thus it appears that in the 
dense limit, the critical temperatures are almost independent of graph 
topology, except for the $d$-dependence. In the sparse limit (see Fig.\
\ref{fig:ndep}.d, note the rescaled vertical axis), the two approximations
show clearly different scaling behavior: the MF
critical $\beta_c^{MF}$ apparently scales like $1/\sqrt{\Delta}$, wherease the
Bethe critical $\beta_c^{MF}$ more closely follows $1/\sqrt{d}$. With
increasing graph size, the difference between both approximation methods
becomes larger and larger. For the BA model, we have not found any theoretical
results concerning phase transitions in the literature, but based on the
analysis for ER graphs and on Monte Carlo experiments we believe that the Bethe
approximation correctly describes the phase transitions.

In conclusion, we have applied the Bethe approximation to spin-glass systems 
on random graphs. We have shown that the Bethe
approximation is more powerful than the naive Mean Field approximation and that
our approach agrees with previous results using replica methods. The advantage
of our approach is that it 
extends to arbitrary graph topologies and
interactions and works for single instances, which is important for
applications. In this light we would like to mention the connection with the
\emph{Belief Propagation} (BP) algorithm \cite{Pearl88}, also known under the
names \emph{Sum-Product algorithm} and \emph{Loopy Belief Propagation}. BP is a
currently very popular algorithm used in diverse applications to minimize the
Bethe free energy \cite{YedidiaFreemanWeiss01, Heskes03}. As illustrated in
Fig.\ \ref{fig:sg_transition}(c), and what we in fact have observed in all our
numerical experiments, the BP algorithm converges only in the paramagnetic
regime. At the transition to a spin-glass phase, the number of iterations
before convergence explodes. Details are beyond the scope of this Letter and
will be explored in a forthcoming article.

A possible generalization of this work would be the incorporation of local
fields $h_i$. Based on (preliminary) numerical experiments, we expect that
adding local fields \emph{increases} the Bethe critical inverse temperature
$\beta_c^{Be}$, thereby extending the paramagnetic regime (it is easy to see
that the Mean Field critical $\beta_c^{MF}$ increases).

\begin{acknowledgments}
This research was supported in part by the Dutch Technology Foundation
(STW). We thank Tom Heskes and Kees Albers for useful suggestions and
interesting discussions.
\end{acknowledgments}

\end{document}